*Cisplatin coordination chemistry determination at hen egg white lysozyme His15 with ligand distances and angles, and their standard uncertainties, and also reporting a split occupancy effect*

John R. Helliwell[1], Simon W. M. Tanley[1], Antoine M. M. Schreurs[2] and Loes M. J. Kroon-Batenburg[2]

1.School of Chemistry, Faculty of Engineering and Physical Sciences, University of Manchester, Brunswick Street, Manchester M13 9PL, England.

2. Crystal and Structural Chemistry, Bijvoet Center for Biomolecular Research, Faculty of Science, Utrecht University, Padualaan 8, 3584 CH Utrecht, The Netherlands.

john.helliwell@manchester.ac.uk

**Abstract**

Following the interest of L Messori and A Merlino 2016 Coordination Chemistry Reviews (2016, 315, 67–89) in the platinum ions' coordination geometries in our PDB entries 4dd4 and 4dd6 we have extended our analyses to find out if the precise geometric details they seek, with distance and angle standard uncertainties, are possible. For the case of 4dd4 it has proved possible and whereby we made a comparison of with-ligand-restraints and without-ligand-restraints models, and derived the standard uncertainties for the latter using the Diffraction Precision Index ('DPI')method. This proved not to be possible for 4dd6-new however due to a split occupancy effect on the bound cisplatin, and whose chemical identity is not certain. It is very interesting what they are though, and for which attention has also been drawn to such an effect in the studies reported by Ferraro et al 2016. We plan then an extensive reanalysis of all the existing datasets of cisplatin bound to histidine with a detailed look for such split occupancy effects, which we will now undertake and report later in the rather new 'data review' style of article. All these extended analyses, reported here and that we plan in our data review, are outside the scope of our original paper (Tanley et al 2012) and whose conclusions are obviously not affected, namely that DMSO is important for cisplatin binding to histidine versus under aqueous conditions. Such new and extended analyses, based on improved crystal structure analysis tools and/or more recent structural insights, are likely to be increasingly important in macromolecular crystallography and are within the vision of the 'Living PDB and Living Publication' (espoused by Helliwell, Terwilliger and McMahon (2012).

**Summary**

We have extended our previously published crystal structure refinements of cisplatin binding to hen egg lysozyme using the existing diffraction data for our PDB entries 4dd4 and 4dd6 (Tanley et al 2012). This follows the interest of Messori and Merlino 2016 in the platinum ions' precise coordination geometry. The chemical formula for the ligands to the platinum ions for 4dd4-new are $[PtClNH_3NH_3His15]$ for $Pt^{\delta}$, and $[PtClNH_3His15Arg14]$ for $Pt^{\varepsilon}$, and which remain the same but the geometry determination is improved. The chemical formula for the ligands to the platinum ions for 4dd6-new also remains the same for $Pt^{\delta}$, which is $[PtClNH_3NH_3His15]$ and $[PtClNH_3His15Arg14]$ for $Pt^{\varepsilon}$ but there is also evidence for a minor fraction split occupancy and for which it is unclear if a (quite possibly) bound water may be involved. The difference, albeit small, of the cisplatin binding mode at $Pt^{\varepsilon}$ for 4dd6-new versus 4dd4-new is interesting because each crystal used for data collection was from the same batch method crystallisation pot but the cryo-protectants for the diffraction data collection were different, respectively glycerol and paratone. The $Pt^{\delta}$ ligand density for 4dd6-new at the position trans to the Cl ligand shows some splitting, again indicating a split occupancy of some kind. Multiple conformers for such cisplatin binding histidine involving histidine and arginine has been noted by Ferraro et al 2016. The final model details at His 15 for 4dd4-new and 4dd6-new are shown in Figure 1.

It was not possible to extend the diffraction data resolution limit via a re- processing of 4dd4 or 4dd6 diffraction data, which is limited simply by the geometry of the measuring protocol used. The diffraction data sets were processed using EVAL15 (Schreurs et al., 2010) as previously described in Tanley et al 2012.

The focus of our first paper Tanley et al 2012 was on the platinum ion occupancies as a function of use of DMSO or not, and whose observations are entirely consistent with these two improved model refinements. As noted in detail in Tanley et al 2012 the anomalous difference peaks on the platinum atoms are highly significant (above 25 sigma in each case, see Figure 1). The PDB depositions for 4dd4-new and 4dd6-new are now replaced by new PDB depositions with codes:- 5L3H and 5L3I.

Overall, the coordination geometry that each $Pt^{\delta}$ shows in 4dd4-new and 4dd6-new, for the [$PtClNH_3NH_3His15$] moiety, are with the ligands in a square planar arrangement. This is the same as for 5hll (Tanley et al 2016) and highly similar to the findings of Ferraro et al 2016. Likewise the coordination geometry that each $Pt^{\varepsilon}$ shows in 4dd4-new and 4dd6-new, for the [$PtClNH_3His15Arg14$] moiety, are with the ligands in a square planar arrangement.

Table S1 provides a summary of the diffraction data, which are unchanged obviously but repeated here for ease of reading, and the final model refinement cycle' details for this extended analysis.

Table S2 provides the precise coordination geometry distances and angles, both with and without platinum ligand distance restraints to usual values, for 4dd4-new. The standard uncertainties for the platinum coordination geometries at His15 were calculated using the Cruickshank (1999) DPI via the 'DPI webserver' for this purpose (Kumar et al 2015) for a final round of model refinement where the 'platinum ligand distance restraints' were removed.

**Acknowledgements**
We thank Messori and Merlino 2016 Coordination Chemistry Reviews (2016) 315, 67–89
for their interest in the precise coordination chemistry geometry of our platinum ion binding sites to His15 in hen egg white lysozyme (see their Figure 7) and which have led to these extended model refinements reported here and quantification of the 4dd4-new platinum ligand distances and angles along with their standard uncertainties. We thank Prof Dr Norbert Strater for detailed discussions.

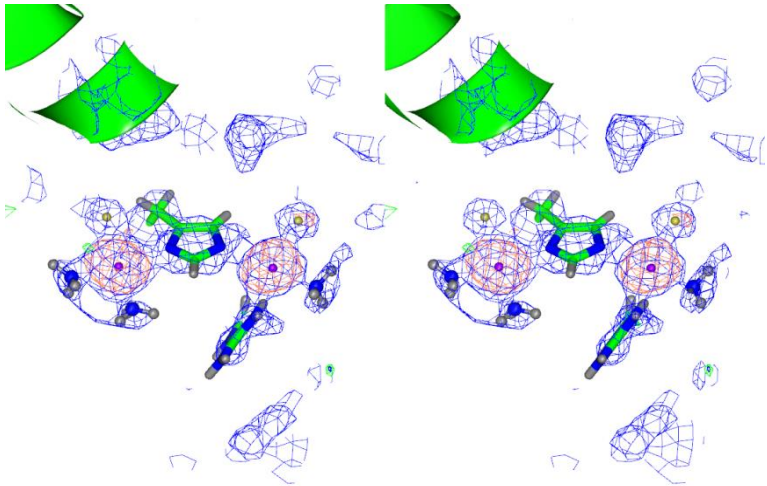
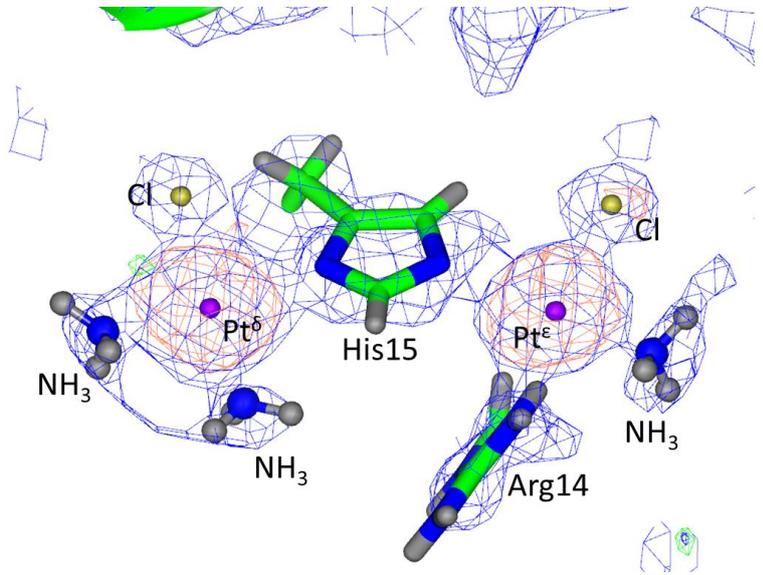
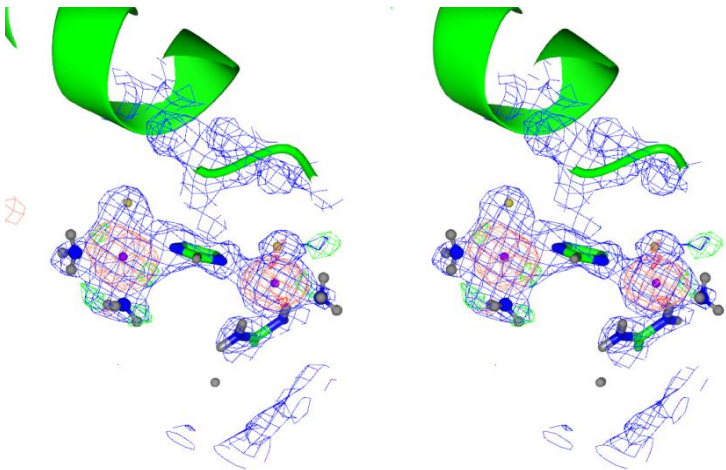

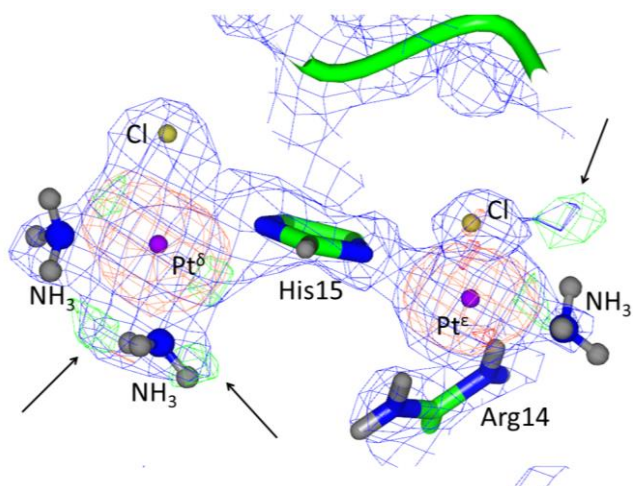

Figure 1 Stereo electron density maps (2Fo-Fc) in blue at 1.2rms and (Fo-Fc) in green at 4.0 σ and the anomalous difference density in orange at 3.5σ at His 15; (top) 4dd4-new model (bottom) 4dd6-new model. As noted in detail in Tanley et al 2012 the anomalous difference peaks on the platinum atoms are highly significant (above 25 sigma in each case). The figure for each presented below the respective stereo figure display shows the atom labels; for 4dd6-new the arrows highlight the split occupancy electron density peaks and whose chemical identity cannot be assigned from this one study but attention to such an effect, which is indeed interesting, has been drawn by Ferraro et al 2016. The radii of the circles used to show the positions of the different atoms are indicative i.e. not meant to be actual atomic or ionic radii. The knowledge of the positions and orientations of the hydrogen atoms at this diffraction resolution are also indicative.

Supplementary

Overall approach

As an extension of our original model refinement (Tanley et al 2012) we have now been able to use the 'validation within model refinement computational environment' of Phenix_Refine (Afonine et al 2010) that is now available within the Windows computer environment, and which we have found very advantageous e.g. for discerning solute molecule assignments and avoiding clashes with the protein. As a second change we have also removed the use of the PDB's CPT ligand descriptor moiety, which has an incorrect tetrahedral geometry and is confusing to the coordinates file user.

4dd4:

The PDB's CPT ligand descriptors were deleted and an omit electron density map calculated. For the His15 ND side of the imidazole ring the omit electron density map showed three peaks at the expected geometric positions for the platinum ligands forming a square planar coordination chemistry geometric arrangement. The omit map peak heights were 9.4σ, 5.6σ and 4.9σ and from which the assignment of chloride and two ammines respectively were made. The His15 NE position cisplatin binding site omit map showed one clear ligand (9.4 σ) and was assigned to a chloride and another of this platinum atom's ligands with an omit peak height of 5.7 σ was assigned to an ammine.  The nitrogen of the Arg14 side chain was clearly the final ligand for the His 15 NE platinum ion position. PHENIX_REFINE (Afonine et al 2012) was used for model refinement; the solute molecules were checked again as per the Critique article of Shabalin et al 2015 and to which we promptly responded (Tanley et al 2015). The model refinement statistics are given in Table S1.

4dd6:

The PDB's CPT ligand descriptors were deleted and an omit electron density map calculated. For the His15 ND side of the imidazole ring the omit electron density map showed three peaks at the expected geometric positions for the platinum ligands forming a square planar coordination chemistry geometric arrangement. The omit map peak heights were 7.8σ, 6.0σ and 5.5σ and from which the assignment of chloride and two ammines respectively were made. The His15 NE position cisplatin binding site omit map showed one clear ligand (7.6σ) and was assigned to a chloride  and another of this platinum atom's ligands with an omit peak height of 5.8 σ was assigned to an ammine. There was an additional nearby difference map peak at 5.3σ, whose chemical identity is not clear but might be a static disorder of some kind (possibly a water molecule?).  The placement of the Arg 14 side chain shows a bidentate arrangement to the PT NE. The PT ND ammine ligand trans to the chlorine also shows some split (i.e. static disorder) electron density. The model refinement statistics are given in Table S1.

Table S1 Data processing and final model re refinement details.

(a) 4dd4 Cycle 18 (new PDB code 5L3H)

| | |
|---|---|
| Wavelength (Å) | 1.54 |
| Temperature of data collection | 100K |
| Resolution range (Å) | 24.93 - 1.7 (1.761 - 1.7) |
| Space group | P 43 21 2 |
| Unit cell | 78.83 78.83 37.02 90 90 90 |
| Total reflections | 173061 |
| Unique reflections | 11826 (1314) |
| Multiplicity | 14.6 (9.1) |
| Completeness (%) | 88.59 (100.00) |
| Mean I/sigma(I) | 22.2 (5.2) |
| Wilson B-factor | 12.70 |
| R-meas | 0.079 (0.313) |
| Reflections used for R-free | 582 |
| R-work | 0.2020 (0.2138) |
| R-free | 0.2476 (0.2839) |
| Number of non-hydrogen atoms | 1149 |
| macromolecules | 1050 |
| ligands | 20 |
| water | 79 |
| RMS(bonds) | 0.003 |
| RMS(angles) | 0.72 |
| Ramachandran favored (%) | 98 |
| Ramachandran allowed (%) | 2 |
| Ramachandran outliers (%) | 0 |
| Clashscore | 0.478 |
| Average B-factor | 12.60 |
| macromolecules | 12.40 |
| ligands | 17.60 |
| solvent | 14.30 |
| Cruickshank 'DPI' (Å) | 0.13 |

(b) 4dd6-new cycle 24 (new PDB code 5L3I)

| | |
|---|---|
| Wavelength (Å) | 1.54 |
| Temperature of data collection (K) | 100 |
| Resolution range (Å) | 18.39 - 1.7 (1.76 - 1.7) |
| Space group | P 43 21 2 |
| Unit cell | 78.02 78.02 37.07 90 90 90 |
| Total reflections | 272733 |
| Unique reflections | 10869 (1280) |
| Multiplicity | 25.0 (18.0) |
| Completeness (%) | 83.05 (100.00) |
| Mean I/sigma(I) | 35.1 (8.1) |
| Wilson B-factor | 13.40 |
| R-meas | 0.067 (0.306) |
| Reflections used for R-free | 530 |
| R-work | 0.1810 (0.1893) |
| R-free | 0.2121 (0.2580) |
| Number of non-hydrogen atoms | 1152 |
| macromolecules | 1064 |
| ligands | 27 |
| water | 61 |
| RMS(bonds) | 0.005 |
| RMS(angles) | 0.99 |
| Ramachandran favored (%) | 99 |
| Ramachandran allowed (%) | 1 |
| Ramachandran outliers (%) | 0 |
| Clashscore | 1.89 |
| Average B-factor | 15.20 |
| macromolecules | 14.80 |
| ligands | 25.20 |
| solvent | 17.60 |
| Cruickshank 'DPI' (Å) | 0.13 |

Table S2 The platinum coordination geometry (a) ligand distances and (b) angles and their standard uncertainties in brackets derived from the Cruickshank dpi calculated using the webserver 'calc dpi' (Kumar et al 2015).

| Ligand $^{\%}$ | 4dd4-new distance Å; ligand restraints applied (5L3H) | 4dd4-new distance Å; ligand restraints removed |
|---|---|---|
| $Pt^{\delta}$ to Cl | 2.3 | 2.3(0.2) |
| $Pt^{\delta}$ to $NH_3$ (end position) nitrogen atom | 2.0 | 2.1(0.2) |
| $Pt^{\delta}$ to $NH_3$ (trans to the Cl) nitrogen atom | 2.1 | 2.1(0.2) |
| $Pt^{\delta}$ to His15ND | 2.3 | 2.3(0.2) |
|  |  |  |
|  |  |  |
| $Pt^{\varepsilon}$ to Cl | 2.3 | 2.3(0.2) |
| $Pt^{\varepsilon}$ to $NH_3$ (end position) nitrogen atom | 2.0 | 2.2(0.2) |
| $Pt^{\varepsilon}$ to Arg14N | 2.2 | 2.3(0.2) |
| $Pt^{\varepsilon}$ to His15 NE | 2.4 | 2.4(0.2) |
|  |  |  |

| Angle (degrees) | 4dd4-new distance Å; ligand restraints applied (5L3H) | 4dd4-new distance Å; ligand restraints removed |
|---|---|---|
| Cl to $Pt^{\delta}$ to $NH_3$ (end position) nitrogen atom | 87 | 87 |
| $NH_3$ (trans to | 98 | 98 |
| the Cl) nitrogen atom to $Pt^{\delta}$ to $NH_3$ (end position) nitrogen atom) | | |
| Cl to $Pt^{\delta}$ to His15ND | 85 | 85 |
| $NH_3$ (trans to the Cl) nitrogen atom to $Pt^{\delta}$ to His15ND | 91 | 91 |
| | | |
| Cl to $Pt^{\varepsilon}$ to $NH_3$ (end position) nitrogen atom) | 83 | 88 |
| Cl to $Pt^{\varepsilon}$ to His15ND | 97 | 96 |
| His15 NE to $Pt^{\varepsilon}$ to Arg14N | 90 | 90 |

**Footnote**

% The platinum atom and associated ligand occupancies were set equal. Their respective refined atomic displacement parameters show a close correspondence (small spread) as noted also by Tanley et al 2016, and being therefore a further validation of correct ligand assignments, as well as the initial omit difference electron density relative peak heights, as described in the text.
& Using the standard uncertainties of the ligand distances in Table S2a and the uncertainty on these angles is approximately 10 degrees.